\documentclass[a4paper,fleqn,usenatbib]{mnras}
\usepackage{graphicx}
\usepackage{psfrag}
\usepackage{amsmath}
\usepackage{amssymb}

\newcommand{\be}{\begin{equation}}
\newcommand{\e}{\end{equation}}
\newcommand{\bear}{\begin{eqnarray}}
\newcommand{\ear}{\end{eqnarray}}

\newcommand{\hmpc}{{\, h^{-1}\, {\rm Mpc}}}
\def\aj{AJ}
\def\apj{ApJ}
\def\apjl{ApJL}
\def\mnras{MNRAS}

\def\aap{A\&A}
\def\apjs{ApJS}
\def\nat{Nature}
\def\physrep{Physics Reports}

\begin{document}

\title[Homogeneity with multiple tracers ]{The many scales to cosmic
  homogeneity: Use of multiple tracers from the SDSS}\author [Sarkar,
  Majumdar, Pandey, Kedia \& Sarkar] {Prakash
  Sarkar$^{1}$\thanks{prakash.sarkar@gmail.com}, Subhabrata
  Majumdar$^2$, Biswajit
  Pandey$^3$\thanks{biswap@visva-bharati.ac.in}, Atul Kedia$^4$
  \newauthor{ and Suman Sarkar$^3$} \\ $^1$ Department of Physics,
  National Institute of Technology, Jamshedpur,831014, India \\ $^2$
  Tata Institute of Fundamental Research, Mumbai, 400005, India\\ $^3$
  Department of Physics, Visva-Bharati University, Santiniketan,
  Birbhum, 731235, India \\ $^4$ Department of Physics, Indian
  Institute of Technology Bombay Powai, Mumbai - 400076, India }

\maketitle

\begin{abstract}
  
  We carry out multifractal analyses of multiple tracers namely the
  main galaxy sample, the LRG sample and the quasar sample from the
  SDSS to test the assumption of cosmic homogeneity and identify the
  scale of transition to homogeneity, if any. We consider the
  behaviour of the scaled number counts and the scaling relations of
  different moments of the galaxy number counts in spheres of varying
  radius $R$ to calculate the spectrum of the Minkowski-Bouligand
  general dimension $D_{q} (R)$ for $-4 \leq q \leq 4$. The present
  analysis provides us the opportunity to study the spectrum of the
  generalized dimension $D_{q}(R)$ for multiple tracers of the cosmic
  density field over a wide range of length scales and allows us to
  confidently test the validity of the assumption of cosmic
  homogeneity. Our analysis indicates that the SDSS main galaxy sample
  is homogeneous on a length scales of $80 \hmpc$ and beyond whereas
  the SDSS quasar sample and the SDSS LRG sample show transition to
  homogeneity on an even larger length scales at $\sim 150 \hmpc$ and
  $\sim 230 \hmpc$ respectively. These differences in the scale of
  homogeneity arise due to the effective mass and redshift scales
  probed by the different tracers in a Universe where structures form
  hierarchically. Our results reaffirm the validity of cosmic
  homogeneity on large scales irrespective of the tracers used and
  strengthens the foundations of the Standard Model of Cosmology.

\end{abstract}

\section{Introduction}
\label{section:intro} 

The Cosmological Principle is one of the most fundamental assumption
in modern cosmology which states that the Universe is statistically
homogeneous and isotropic on sufficiently large scales. Homogeneity
implies translational invariance i.e. the Universe looks same to all
observers at different locations. Isotropy implies rotational
invariance i.e. the Universe looks the same in all directions. These
two assumptions play an important role in the analysis and
interpretation of various cosmological observations. One can not prove
these assumptions in a rigorously mathematical sense but it is
possible to verify them using various observations. The near uniform
temperature of the cosmic microwave background radiation is considered
to be one of the best possible evidence in favour of isotropy
\citep{penzias,smoot,fixsen}. There are many other observations in
favour of isotropy such as the angular distributions of radio sources
\citep{wilson,blake}, the X-ray background \citep{peeb93,wu,scharf},
the Gamma-ray bursts \citep{meegan,briggs}, the distribution of
galaxies \citep{marinoni,alonso}, the distribution of supernovae
\citep{gupta,lin} and the distribution of neutral hydrogen
\citep{hazra}. But isotropy around us alone does not guarantee
homogeneity of the Universe because homogeneity and isotropy may or
may not coexist. The Universe can be isotropic around a point without
being homogeneous. The homogeneity can be inferred from isotropy only
when there is isotropy around each and every point in the Universe. So
it is not quite straightforward to infer homogeneity of the Universe
from isotropy around us alone.

The statistical properties of galaxy distributions can be
characterized by the correlation functions \citep{peeb80}. The two
point correlation function quantifies the clustering strength which is
well described by a power law on small scales and vanishes on large
scales conforming to large scale homogeneity. However a major drawback
of the correlation function for the analysis of homogeneity arise from
the fact that it assumes a mean density on the scale of the survey
which is not a defined quantity below the scale of homogeneity. It is
well known that the galaxy distribution behave like a fractal on small
scales. But \citet{pietronero} analyzed the CfA1 survey and suggested
that the distribution of galaxies is fractal to arbitrary
large-scales. Further, \citet{coleman92} supported the argument of
Fractal Universe by analyzing different samples. There are other
studies which claim the absence of any transition to homogeneity out
to scale of the survey \citep{amen,joyce,labini07, labini09a,
  labini09b, labini09c, labini11, park16}. \citet{borgani95} claimed
that the fractal structure is valid at small scale but at large-scale
the Universe is Homogeneous. \citet{guzzo97} analyzed the
Perseus-Pisces redshift survey and claimed that galaxies are clustered
at small scale and intermediate scale but is homogeneous on
large-scale. \citet{cappi} found the volume-limited subsample of SSRS2
to be consistent with both fractality at small scale and homogeneity
on large-scale. \citet{bharad99} find homogeneity at scale larger than
80 $\hmpc$, using Multifractal analysis of the LCRS
sample. \citet{pan2000} used fractal analysis to PSCz and find the
homogeneity scale even at 30 $\hmpc$. \citet{yadav} have analyzed the
two-dimensional strips from SDSS DR1 using Multifractal analysis and
found the transition to homogeneity occurring at length-scale $60-70$
$\hmpc$. \citet{hogg} have analyzed the distribution of SDSS Luminous
Red Galaxies (LRG) to find the transition to homogeneity at $\sim 70$
$\hmpc$. \citet{prakash} used the multifractal analysis of SDSS DR6 to
find the transition to homogeneity at length-scale greater than $70$
$\hmpc$. \citet{scrim} find homogeneity at length-scale $70-80$
$\hmpc$ in the redshift range $0.2-0.8$ using WiggleZ survey.
\citet{nadathur} find homogeneity at length-scale above $130$ $\hmpc$
for the SDSS DR7 quasar catalogue containing the Huge-LQG at redshift
$z \sim 1.3$. Some recent studies with Shannon entropy show that the
main galaxies from SDSS DR12 and luminous red galaxies (LRG) from SDSS
DR7 are homogeneous beyond $150 \hmpc$ \citep{pandey15, pandey16}.

As there is no consensus in the issue of cosmic homogeneity, it is
important to test the assumption on multiple datasets with different
statistical tools. If the assumption is ruled out with high
statistical significance by multiple datasets there would be a major
paradigm shift in cosmology.

SDSS is the largest and finest galaxy redshift survey todate which
provides us an unique opportunity to test the assumption of cosmic
homogeneity on very large scales with unprecedented confidence. SDSS
provides distributions of different types of galaxies out to different
distances covering enormous volumes. Galaxies are believed to be a
biased tracer of the underlying mass distribution and different types
of galaxies are expected to have different values of linear bias. In
the present work we would like to test the assumption of homogeneity
using different tracers of the underlying mass distribution. We use
the multifractal analysis \citep{martinez90, coleman92, borgani95} to
characterize the scale of homogeneity in the main galaxy sample, LRG
sample and Quasar sample from the Sloan Digital Sky Survey (SDSS)
\citep{york}. We have considered the main galaxy sample (MGS) from
SDSS DR7, Luminous Red Galaxy (LRG) sample from SDSS DR7 and Quasars
sample from SDSS DR12. The LRG sample covers a much larger volume as
compared to the main galaxy sample but it is only quasi volume limited
and quite sparse as compared to the main sample. The quasar sample
analyzed here is even larger in volume. It is the largest among these
samples but also sparsest among them. These samples provide us the
scope to test homogeneity upto different length scales due to their
different volume coverages. Besides the different volumes and number
densities, the samples also have different properties such as
luminosity and colour resulting in different clustering properties. A
combined analysis of the multiple tracers of the underlying mass
distribution using the same statistical tools, provide us an unique
opportunity to test the assumption of cosmic homogeneity for different
tracers on a wide range of length scales.

A brief outline of the paper follows. In section 2 we describe the
data followed by the method of analysis in section 3. We present the
Results in section 4 and Conclusions in section 5.

\section{Data}

The Sloan Digital Sky Survey (SDSS) \citep{york} is a wide-field
photometric and spectroscopic survey of sky using a $2.5$ m Sloan
Telescope at Apache Point Observatory in New Mexico, United States. It
imaged the sky in five different pass-bands $u$, $g$, $r$, $i$ and
$z$. We have considered three different types of tracers namely the
galaxies from the MAIN sample, the luminous red galaxies (LRGs) and
the quasars from the SDSS. We have used $\Omega_m = 0.31$,
$\Omega_\Lambda = 0.69$ and $h = 0.71$ throughout the analysis. The
various data sets used in the present analysis are described below.

\subsection{SDSS DR7 Main Galaxy sample}

We have used the main galaxy sample from SDSS Data Release 7
\citep{abaz} for which the target selection algorithm is described in
\citet{strauss}. The main galaxy sample comprises of galaxies brighter
than limiting r-band Petrosian magnitude $17.77$. We have identified a
contiguous region in the Northern Galactic cap which spans $-40^\circ
\le\lambda \le 35^\circ$ and $-30^\circ \le \eta \le 30^\circ$, where
$\lambda$ and $\eta$ are survey coordinates defined in
\citet{stout}. We have constructed a volume limited sample of galaxies
in the region by restricting the r-band absolute magnitude to the
range $M_r\le -20$. This absolute magnitude cut produces a volume
limited sample of galaxies within $z\le 0.106$. We finally extract
$64109$ galaxies in the redshift range of $0.040\le z \le 0.106$ which
corresponds to the radial comoving distances in the range $175.24\le r
\le 457.09 \hmpc$ .

\subsection{SDSS DR7 Luminous Red Galaxy sample}

The Luminous Red Galaxy distribution (LRG) \citep{eisen} from the SDSS
extends to a much deeper region of the Universe as compared to the
SDSS Main galaxy sample as they can be observed to greater distances
as compared to normal galaxies for a given magnitude limit. The LRGs
have stable colors which make them relatively easy to pick out from
the rest of the galaxies using the SDSS multi-band photometry. For the
present analysis, we consider the spectroscopic sample of LRG
extracted form the SDSS DR7. The LRG sample is quasi-volume limited to
a redshift of $z=0.36$. We use the DR7-Dim sample extracted by
\citet{kazin}. This sample contains $67,567$ galaxies within the
redshift range of $0.16$ to $0.36$ for g-band absolute Magnitude
($M_g$) range brighter than $-21.2$ and fainter than $-23.2$. We
restrict our sample to a contiguous region in the Northern Galactic
Cap which spans $-52^\circ \le \lambda \le 52^\circ$ and $-31^\circ
\le \eta \le 31^\circ$ which constitutes our sample with $48,308$
LRGs.

\subsection{SDSS DR12 Quasars sample}
The quasars are the brightest class of objects known in the Universe
and their high luminosities allow us to detect them out to larger
distances. We use the SDSS DR12 quasar catalogue for which the target
selection is described in \citet{rossq} and the data is described in
\citet{paris}. The SDSS DR12 catalogue contain a total $297301$
quasars. In the present work we use a quasar sample prepared by
\citet{sarkar}. The preparation of the primary quasar sample is
described in \citet{sarkar}. The primary quasar sample used in this
analysis contain $117882$ quasars with the g-band PSF magnitude $g\leq
22$ and the r-band PSF magnitude $\leq 21.85$ \citep{ross}. We select
the quasars in a contiguous region $150^{\circ} \leq \alpha \leq
240^{\circ}$ and $0^{\circ} \leq \delta \leq 60^{\circ}$ where
$\alpha$ and $\delta$ are the right ascension and declination
respectively. The selected quasars have redshift in the range $2.2
\leq z \leq 3.2$. The resulting quasar sample have a varying number
density in redshift. Constructing a strictly volume limited sample of
quasars from this data results into a sample with a very poor number
density. We construct a quasar sample with near uniform and reasonable
number density in the entire redshift range by using a redshift
dependent magnitude cut $M_{i} \geq M_{lim}(z)$ in the i-band absolute
magnitude. The limiting magnitude $M_{lim}(z)$ is described by a
polynomial of the form $M_{lim}(z)= a z^{3} + b z^{2} + c z + d$ where
$a,b,c$ and $d$ are the coefficients to be determined. We constrain
$a,b,c,d$ so as to have a $<30\%$ fluctuations in the comoving number
density around the mean density at all the redshifts in the chosen
redshift range. We find that $a=24.206, b=-194.325, c=511.413,
d=-467.422$ provides us with a quasar sample consisting of $24,213$
quasars distributed across a volume of $1.83 \times 10^{10}$
${\hmpc}^3$ with a mean number density of $1.32 \times 10^{-6}
h^{3}\,\rm{Mpc}^{-3}$. The quasar sample has a linear extent of
$759.32 \hmpc$ in the radial direction.

\subsection{LasDamas simulations}

Large Suite of Dark Matter Simulations (LasDamas) (MCBride et. al., in
preparation) is a cosmological N-body dark matter simulation project
to obtain adequate resolutions in many large boxes rather than one
single realization at high resolution. The cosmological parameters
used in the LasDamas simulations are those of $\Lambda$CDM model:
$\Omega_m = 0.25$, $\Omega_\Lambda=0.75$, $\Omega_b = 0.04$, $h=0.7$,
$\sigma_8=0.8$ and $n_s=1$. The galaxy mocks are generated by
artificially placing galaxies inside dark matter halos using a halo
occupation distribution with parameters fit from the respective SDSS
galaxy samples. We have used the gamma release of the galaxy mock
sample generated from Oriana simulations. The simulation was carried
out in a simulation box of side $2400 \, \hmpc$ with $1280^3$
particles. The individual particles in the simulation has a mass of
$45.73 \times 10^{10} \, h^{-1} {\rm M}_\odot$ and the spatial
resolution of the simulation is $53$ $h^{-1} \, {\rm Kpc}$. We have
used $30$ and $40$ independent realizations of mock galaxy catalogue
from LasDamas Simulations for the SDSS MAIN and LRG sample
respectively. We then extract mock samples for each of our SDSS MAIN
sample and LRG sample which have the same geometry and number density
as the actual data. The mocks were analyzed exactly in the same way as
the SDSS data.

\begin{figure}
  \includegraphics[height=0.35\textwidth, angle=0]{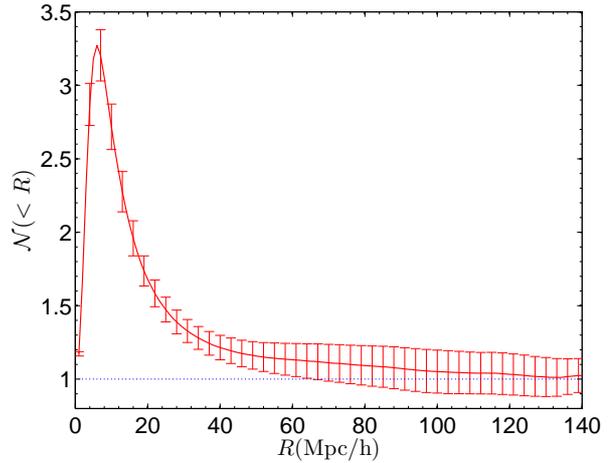}
  \caption{This plot shows the variation of scaled number counts
    $\mathcal{N(<R)}$ with radius $R$ for the SDSS Main Galaxy
    sample. The $1-\sigma$ errorbars are obtained from $30$ Mock
    samples.}
  \label{fig:NvsRMGS}
\end{figure}

\begin{figure*}
 \includegraphics[width=0.45\textwidth]{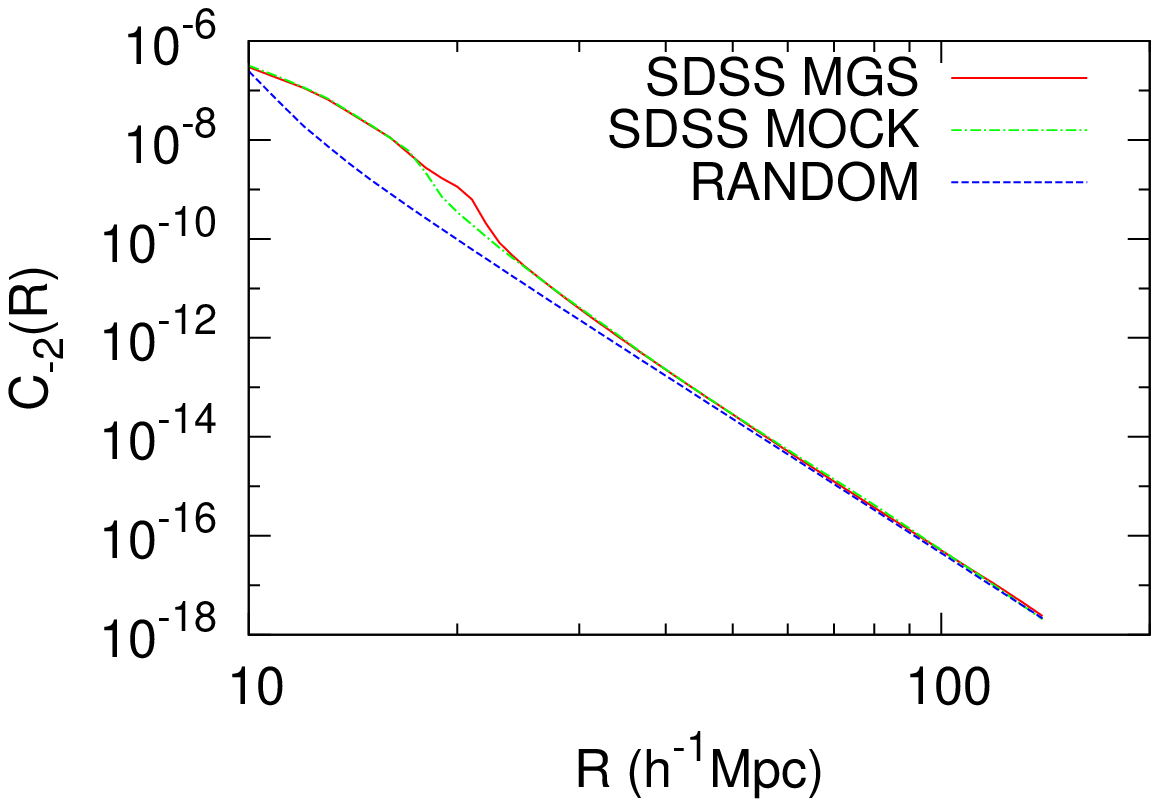}\label{fig:mgs_t1}
  \includegraphics[width=0.45\textwidth]{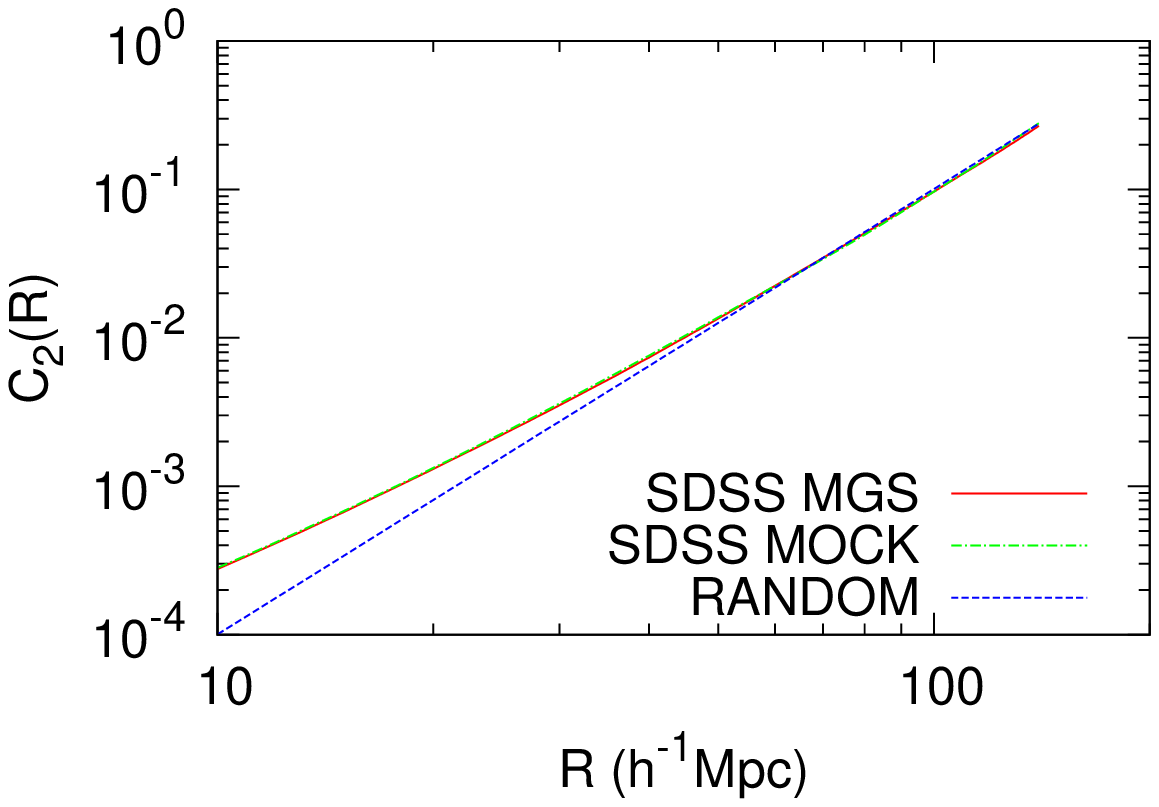}\label{fig:mgs_t2}
  \includegraphics[height=0.45\textwidth,angle=-90]{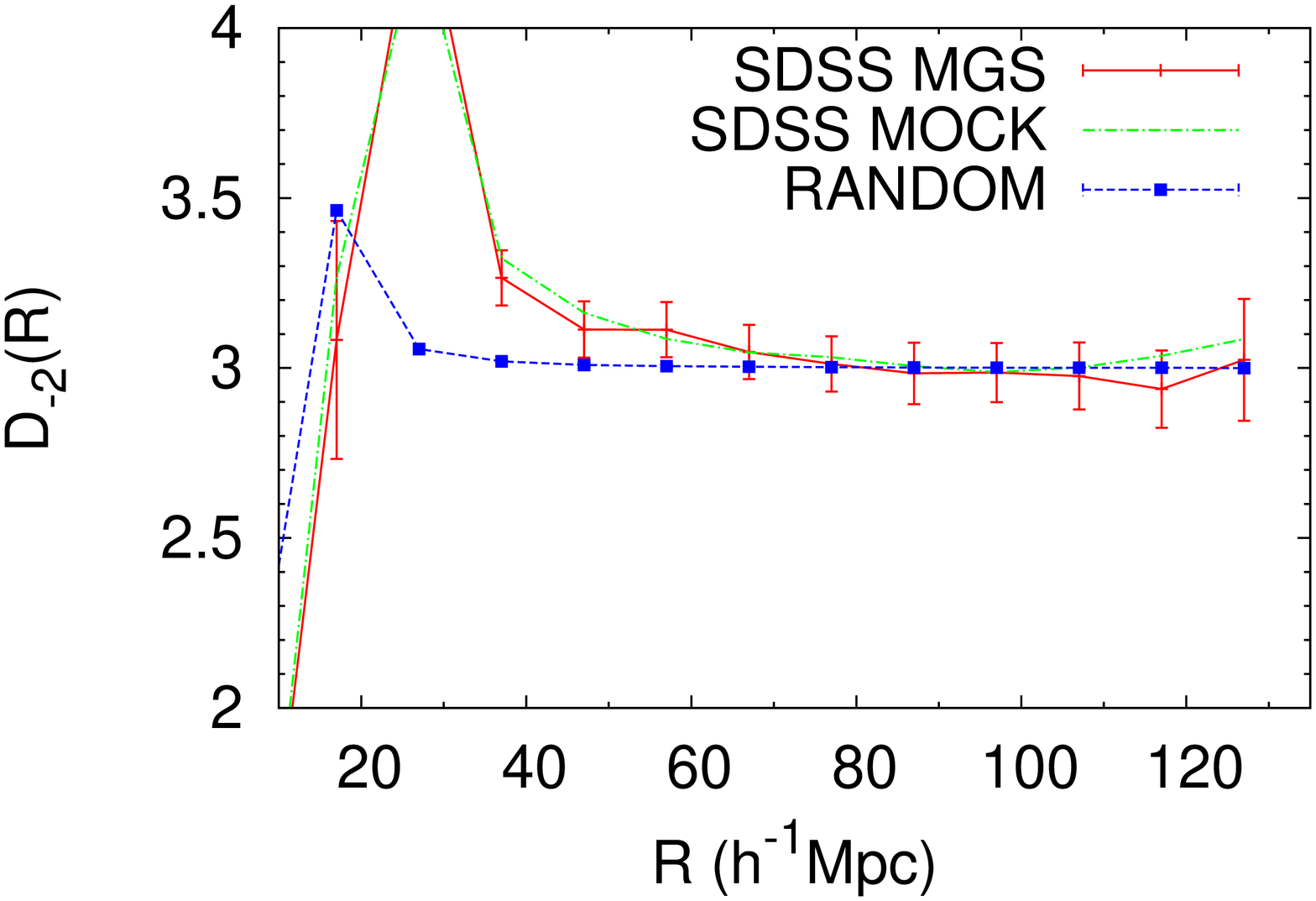}\label{fig:mgs_m1}
  \includegraphics[height=0.45\textwidth, angle=-90]{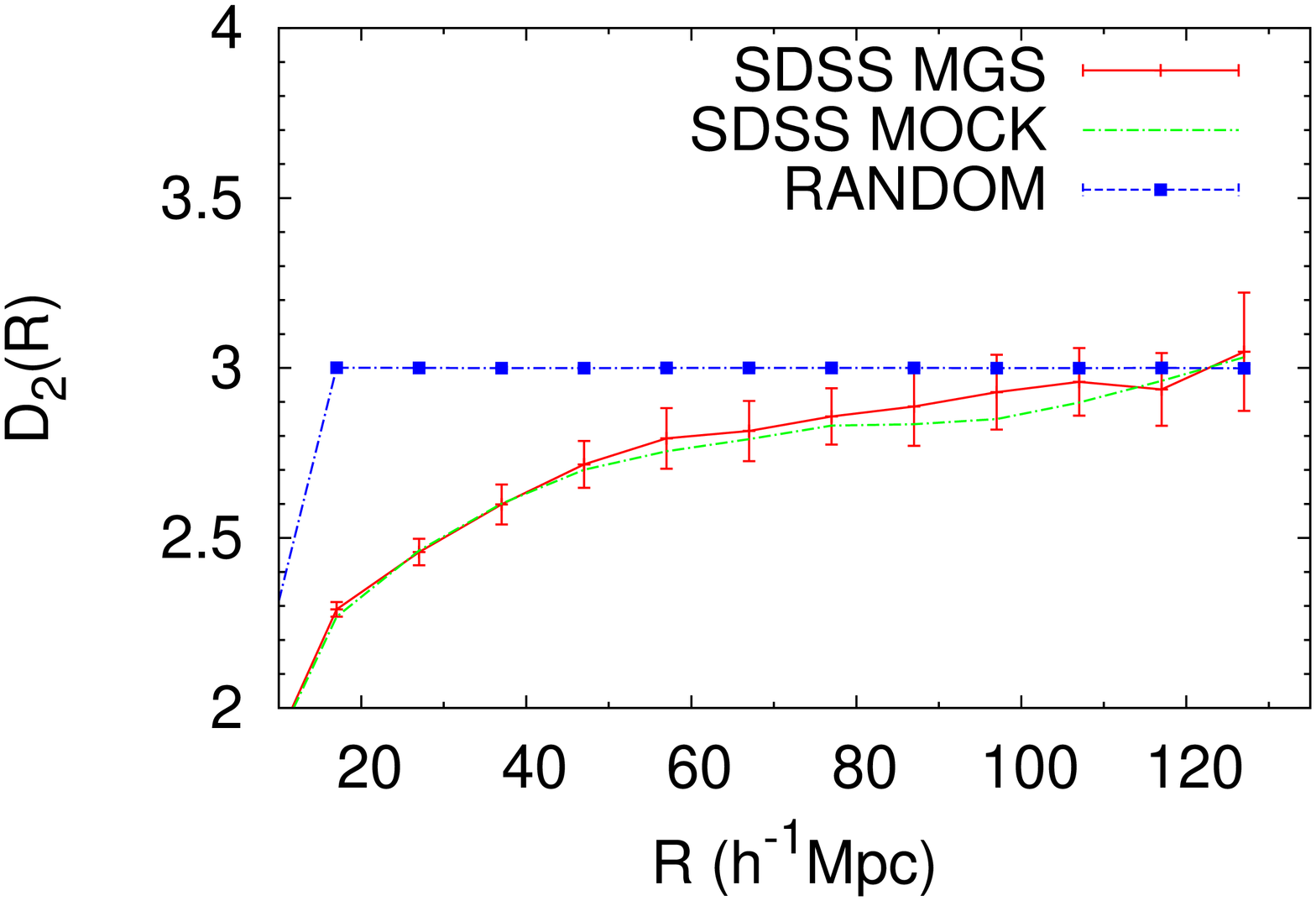}\label{fig:mgs_m2}
  \includegraphics[height=0.45\textwidth, angle=-90]{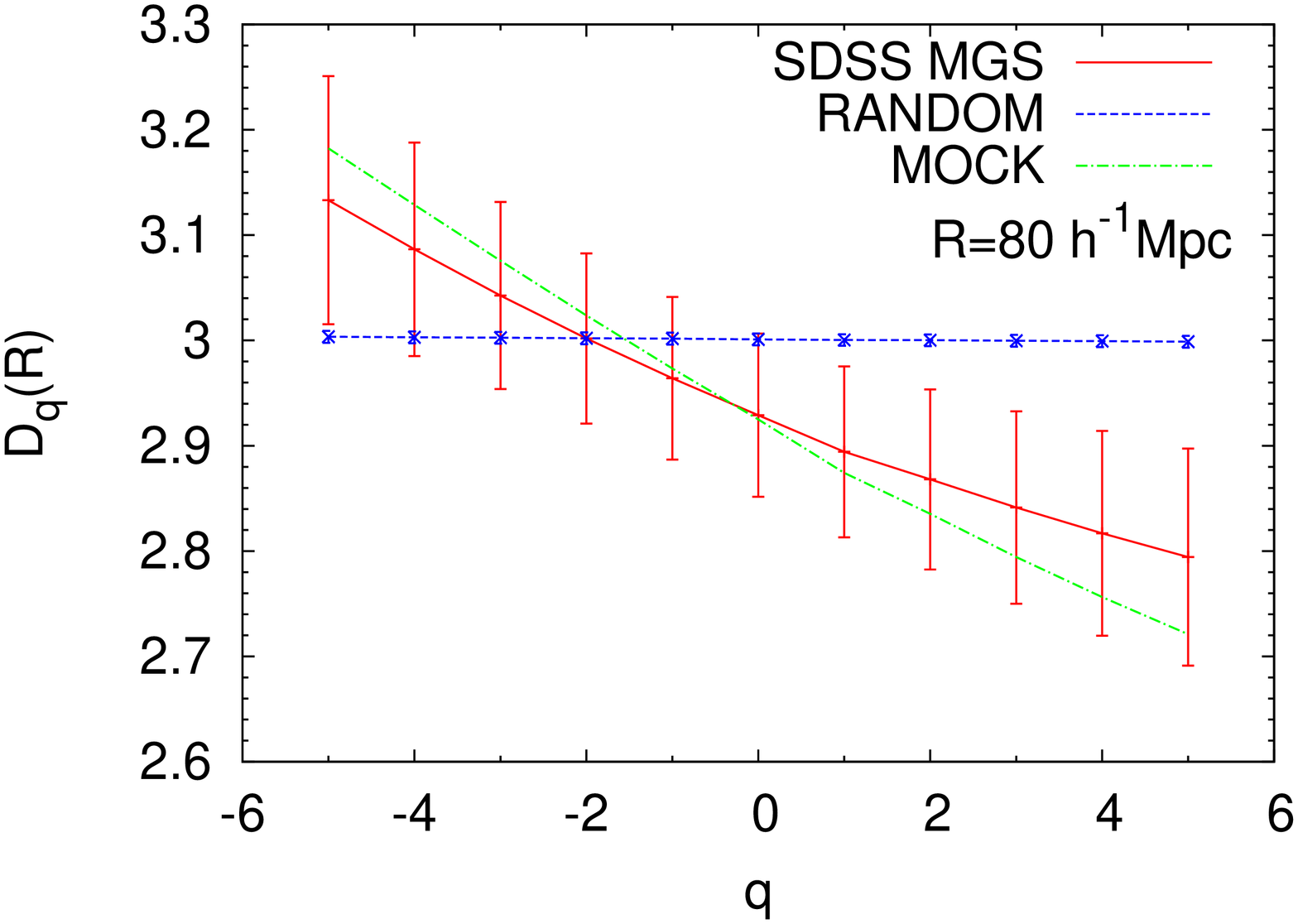}\label{fig:mgs_b1}
  \includegraphics[height=0.45\textwidth, angle=-90]{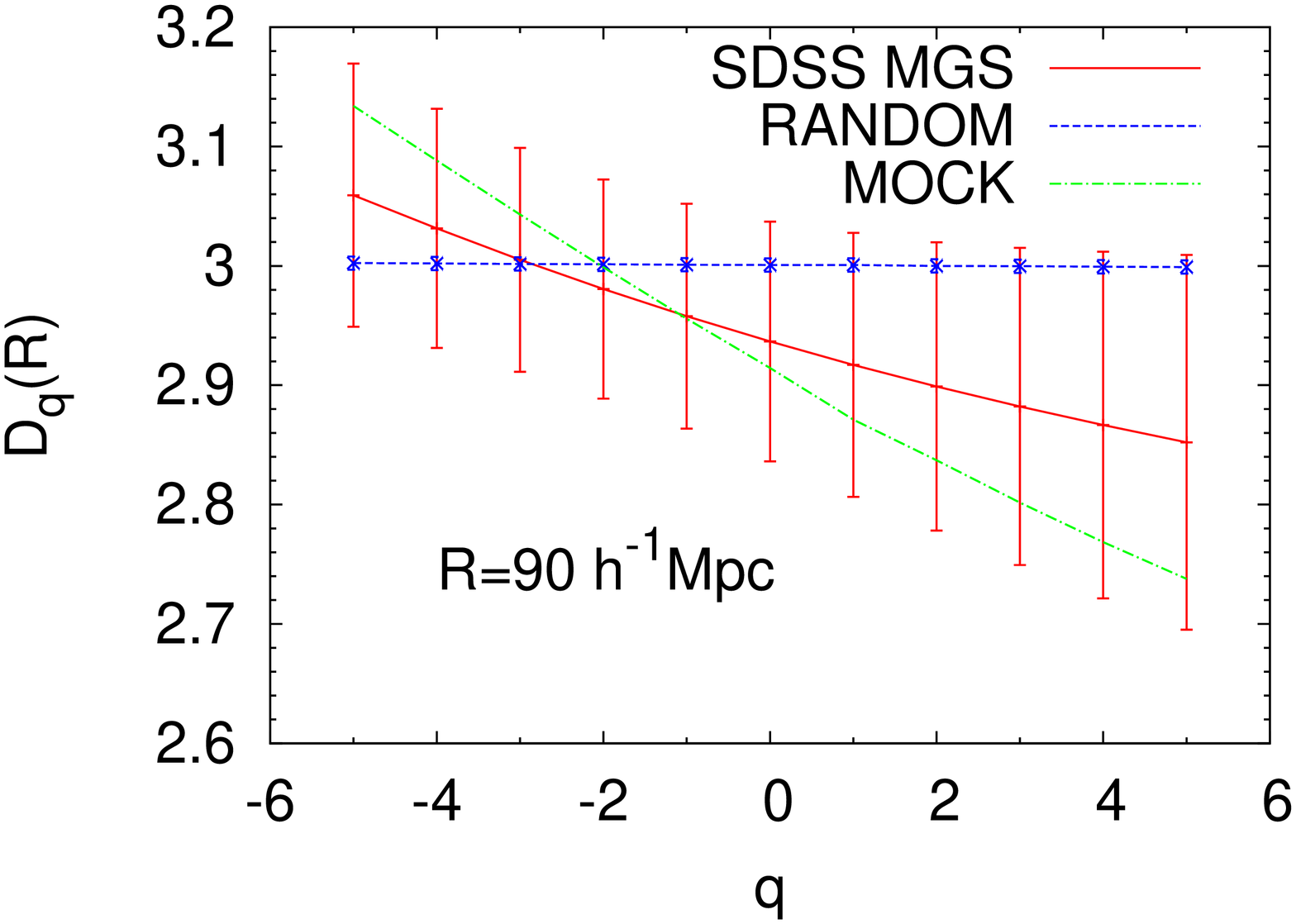}\label{fig:mgs_b2}

\caption{ The top two panels shows the variation of Correlation
  integral $C_q(R)$ with radius $R$ for $q=2$ and $q=-2$ for the SDSS
  main galaxy sample. The variation of $D_q(R)$ with respect to $R$
  for $q=2$ and $q=-2$ for the same galaxy sample are shown in the
  middle two panels. The bottom two panels show the variation of $D_q$
  with $q$ for two different values of $R$ marking the transition to
  homogeneity. The $1-\sigma$ error bars shown in different panels are
  obtained from the mock samples. }
  \label{fig:mgs_all}
\end{figure*}

\begin{figure}
  \includegraphics[height=0.35\textwidth, angle=0]{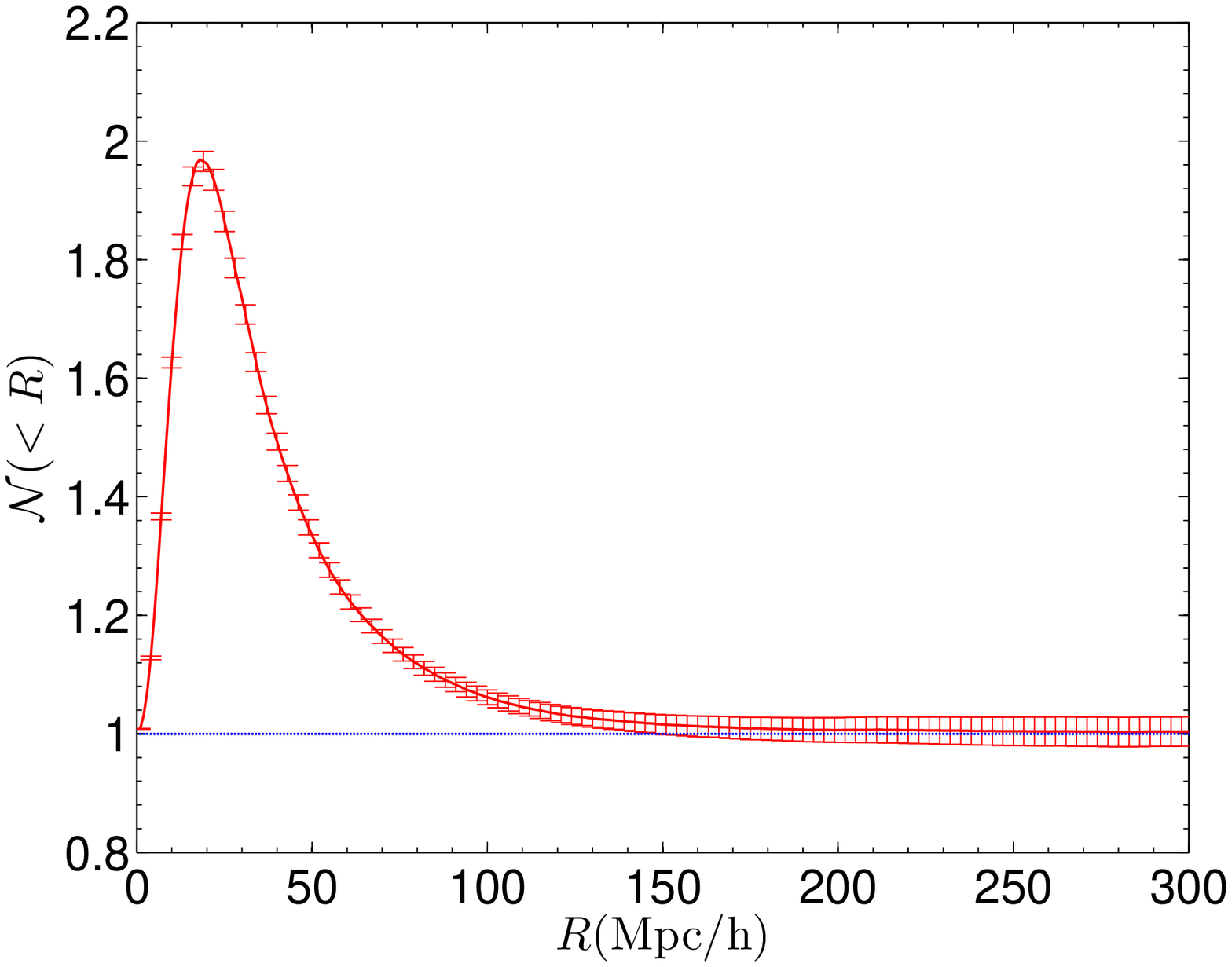} 
  \caption{This plot shows the variation of scaled number counts
    $\mathcal{N(<R)}$ with radius $R$ for the SDSS LRG sample. The
    $1-\sigma$ errorbars are obtained from 40 Mock samples.}
  \label{fig:NvsRLRG}
\end{figure}

\begin{figure*}
  \includegraphics[width=0.45\textwidth]{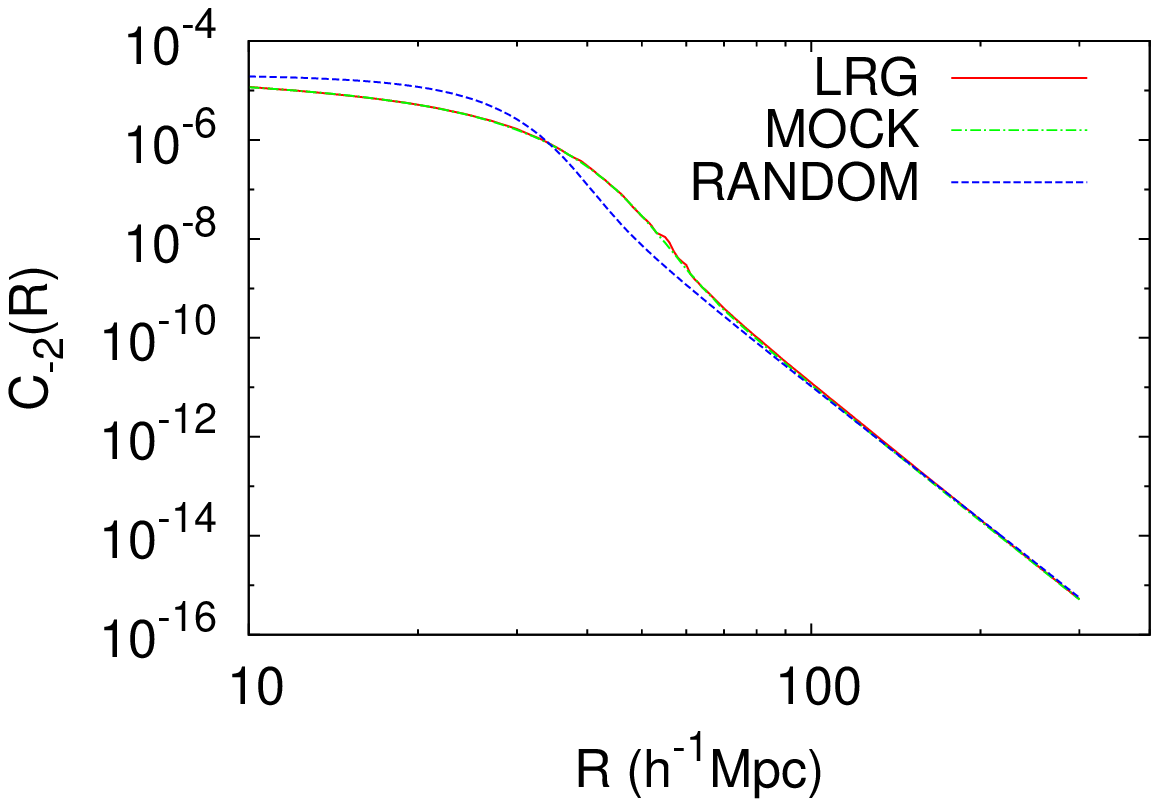}\label{fig:lrg_t1}
  \includegraphics[width=0.45\textwidth]{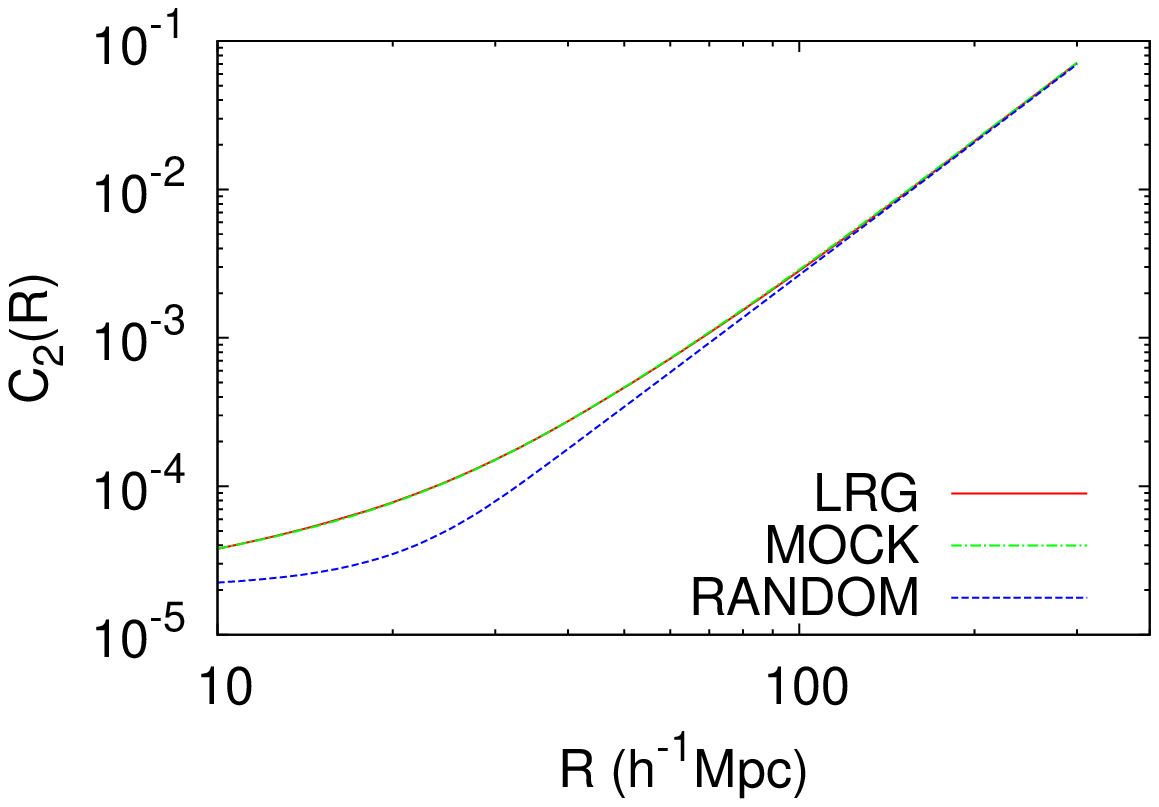}\label{fig:lrg_t2}
  \includegraphics[height=0.45\textwidth, angle=-90]{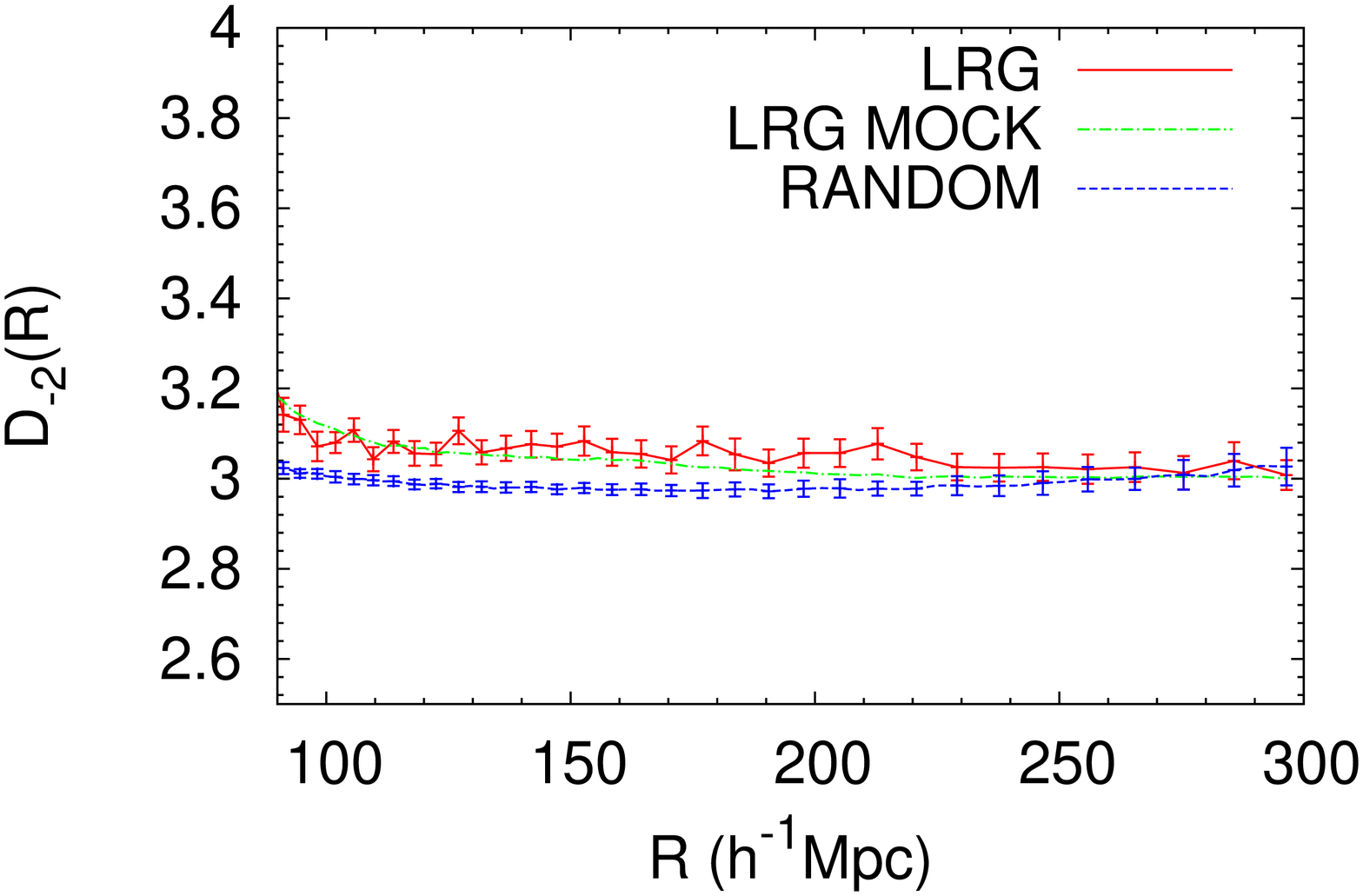}\label{fig:lrg_m1}
  \includegraphics[height=0.45\textwidth, angle=-90]{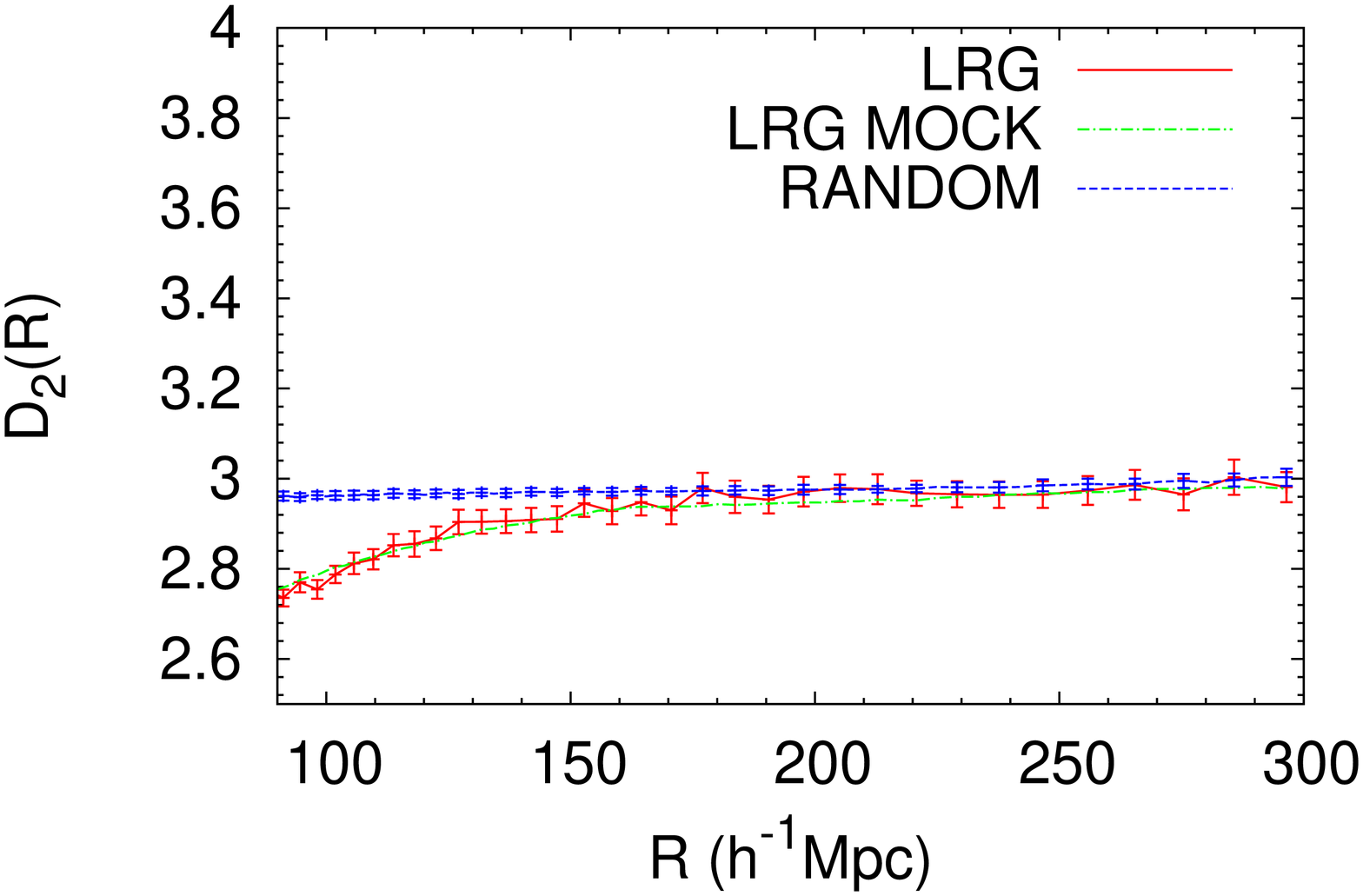}\label{fig:lrg_m2}
  \includegraphics[height=0.45\textwidth, angle=-90]{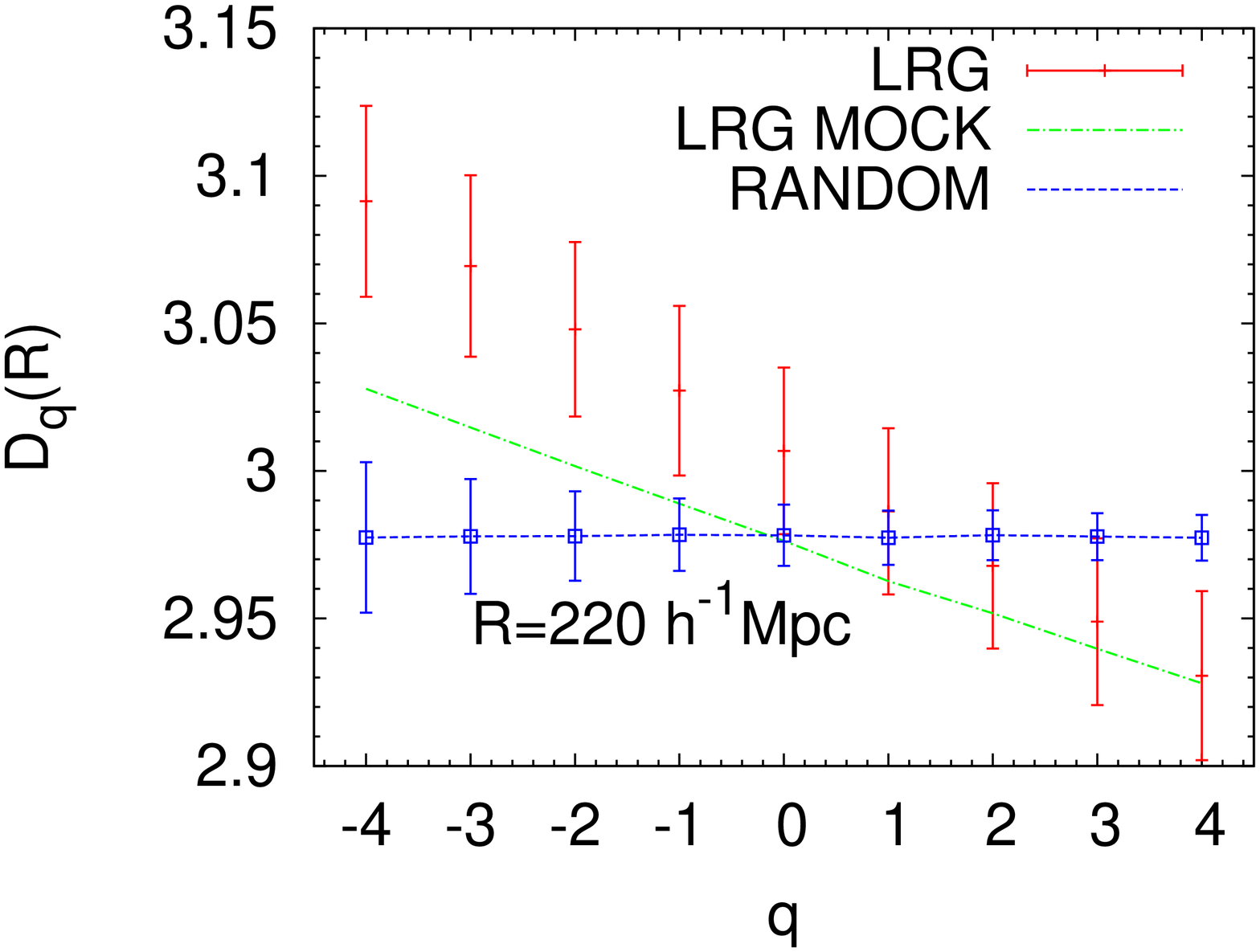}\label{fig:lrg_b1}
  \includegraphics[height=0.45\textwidth, angle=-90]{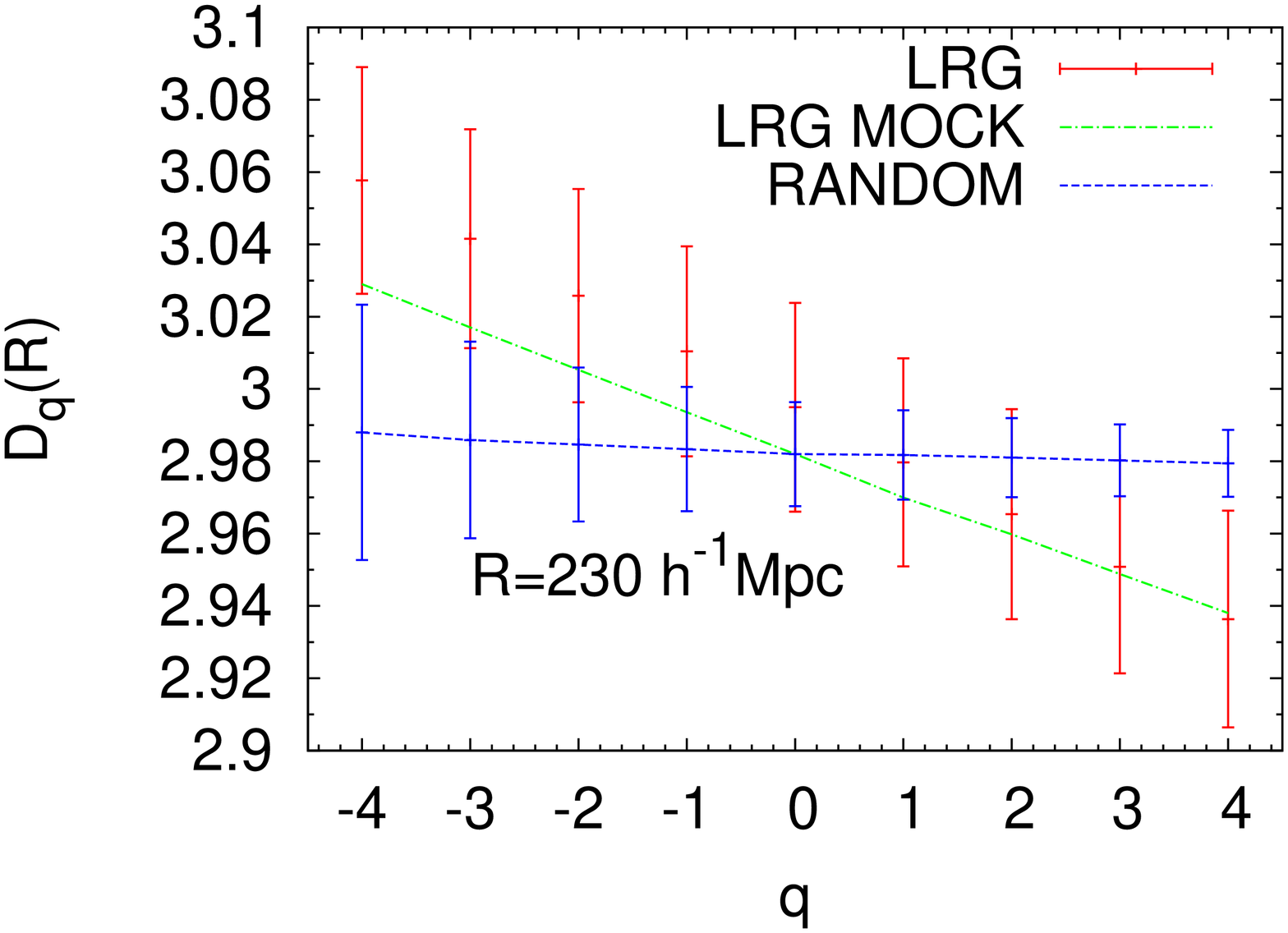}\label{fig:lrg_b2}
\caption{ Same as Figure 2. but for the SDSS LRG sample.
}
\label{fig:lrg_all}
\end{figure*}

\begin{figure}
  \includegraphics[height=0.35\textwidth, angle=0]{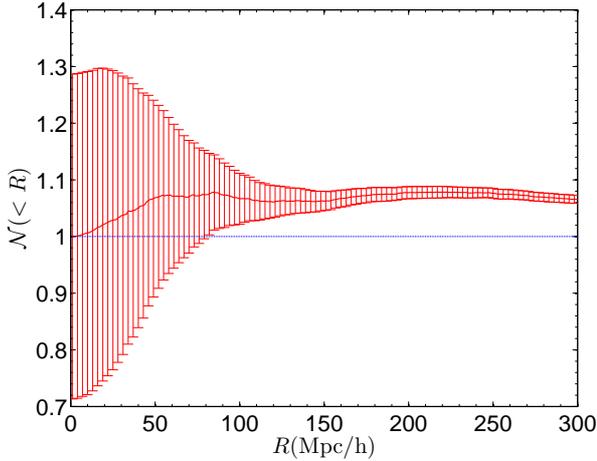}
  \caption{This plot shows the variation of scaled number counts
    $\mathcal{N(<R)}$ with radius $R$ for the SDSS quasar sample. The
    $1-\sigma$ errorbars are obtained from bootstrap resampling.}
  \label{fig:NvsRQSO}
\end{figure}

\begin{figure*}
 \includegraphics[width=0.45\textwidth]{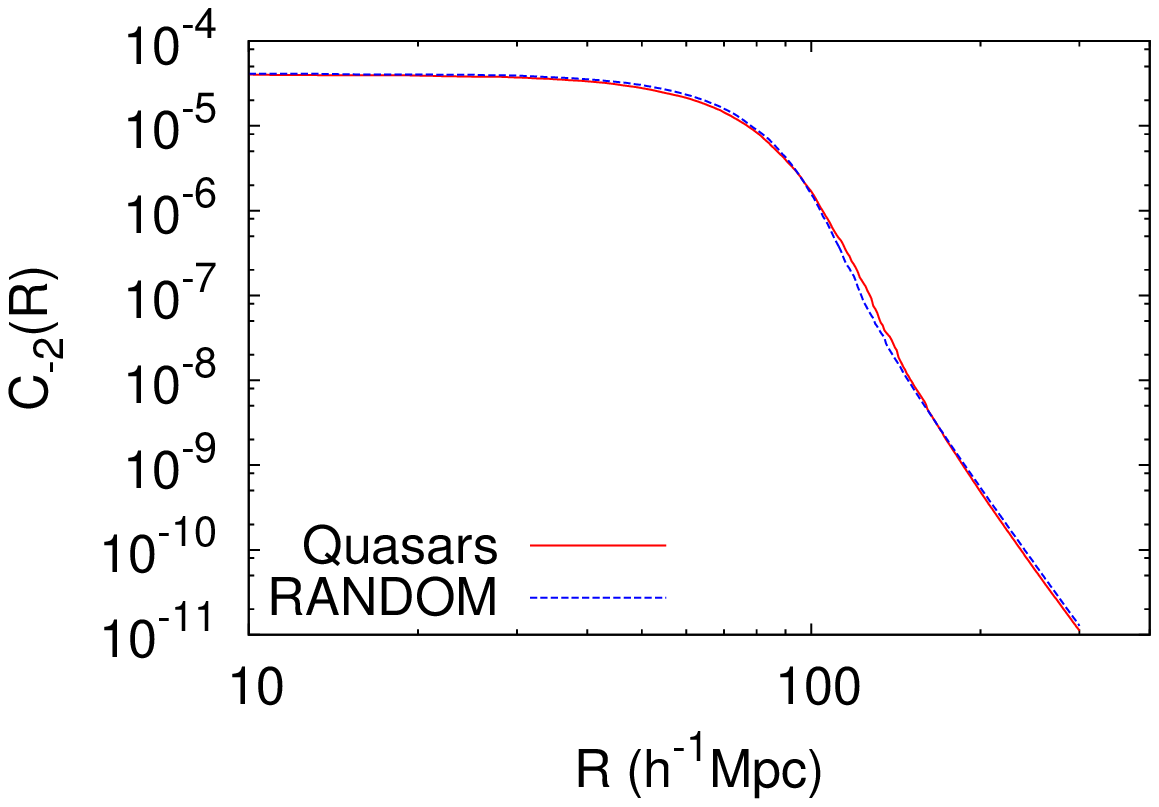}\label{fig:qso_t1}
 \includegraphics[width=0.45\textwidth]{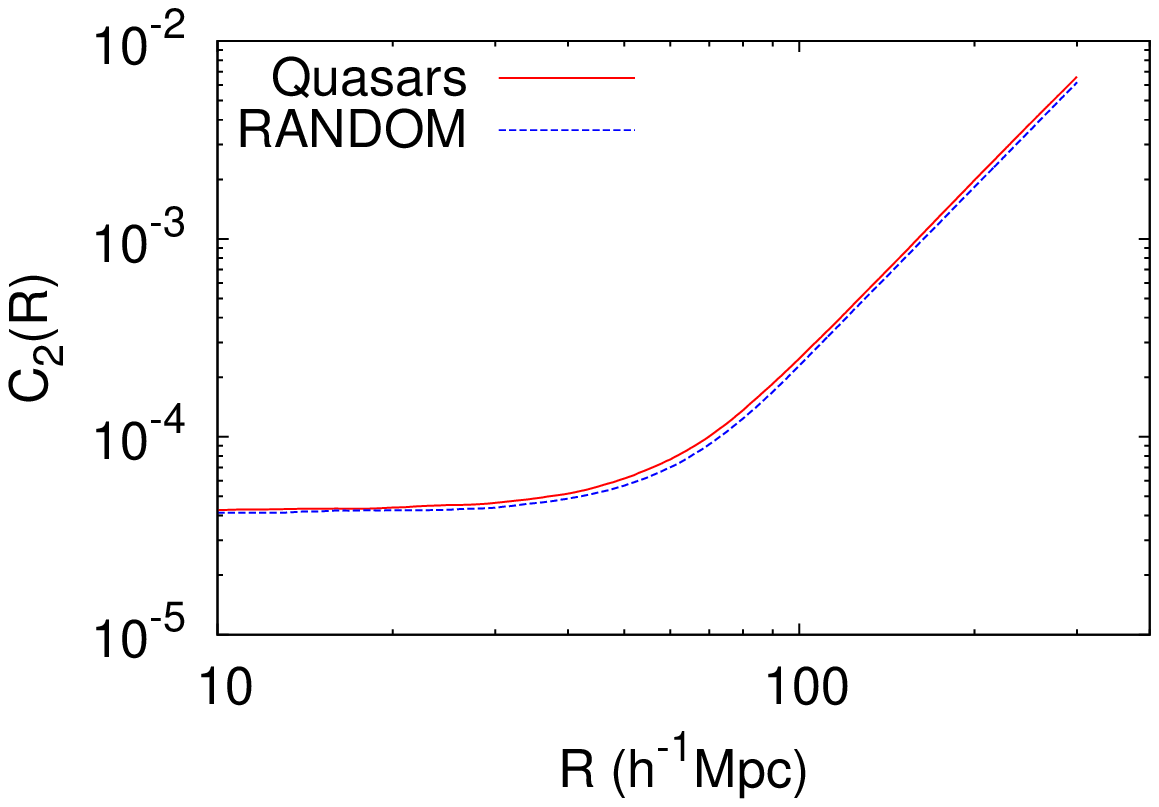}\label{fig:qso_t2}
  \includegraphics[height=0.45\textwidth, angle=-90]{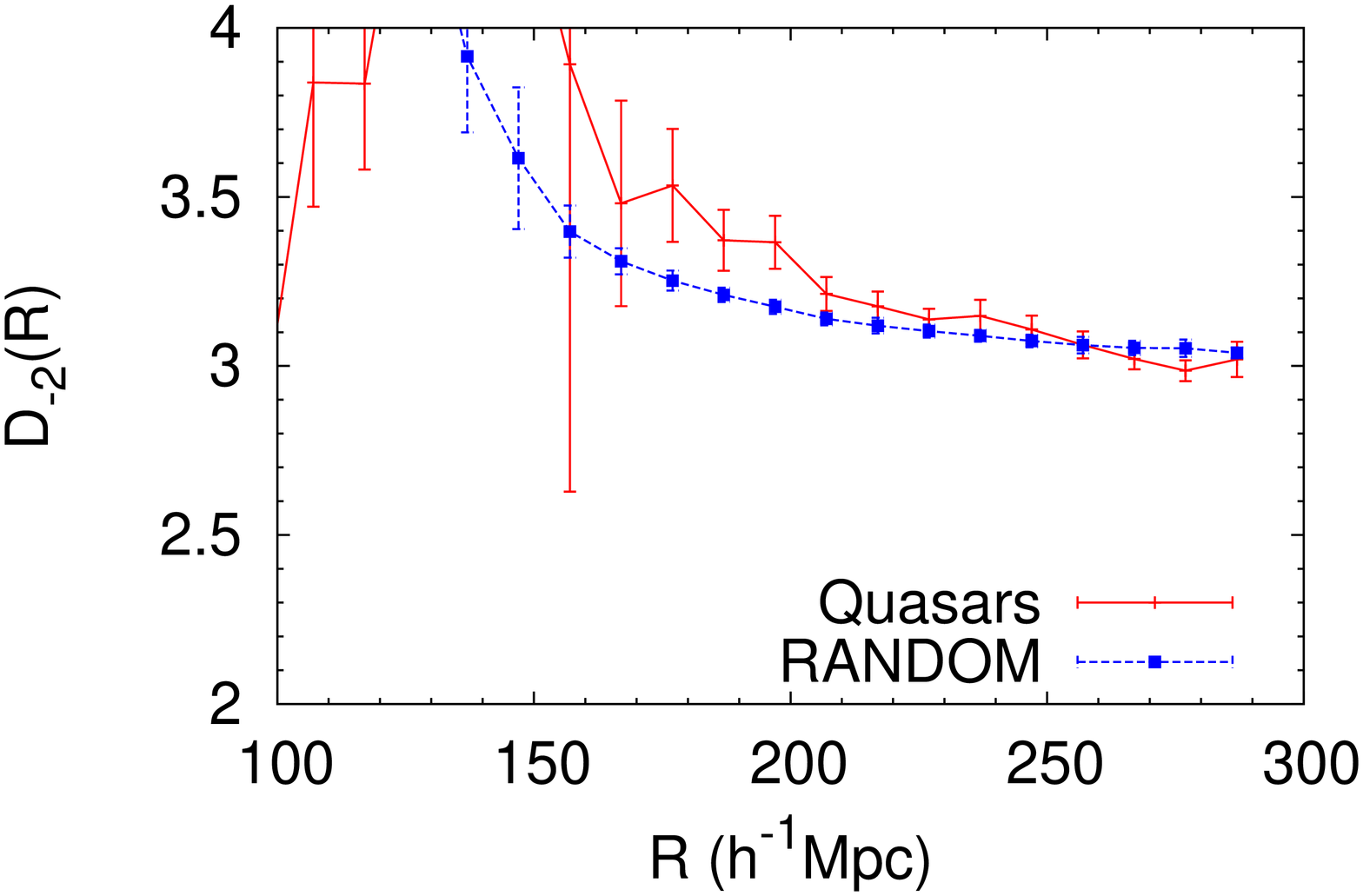}\label{fig:qso_m1}
  \includegraphics[height=0.45\textwidth, angle=-90]{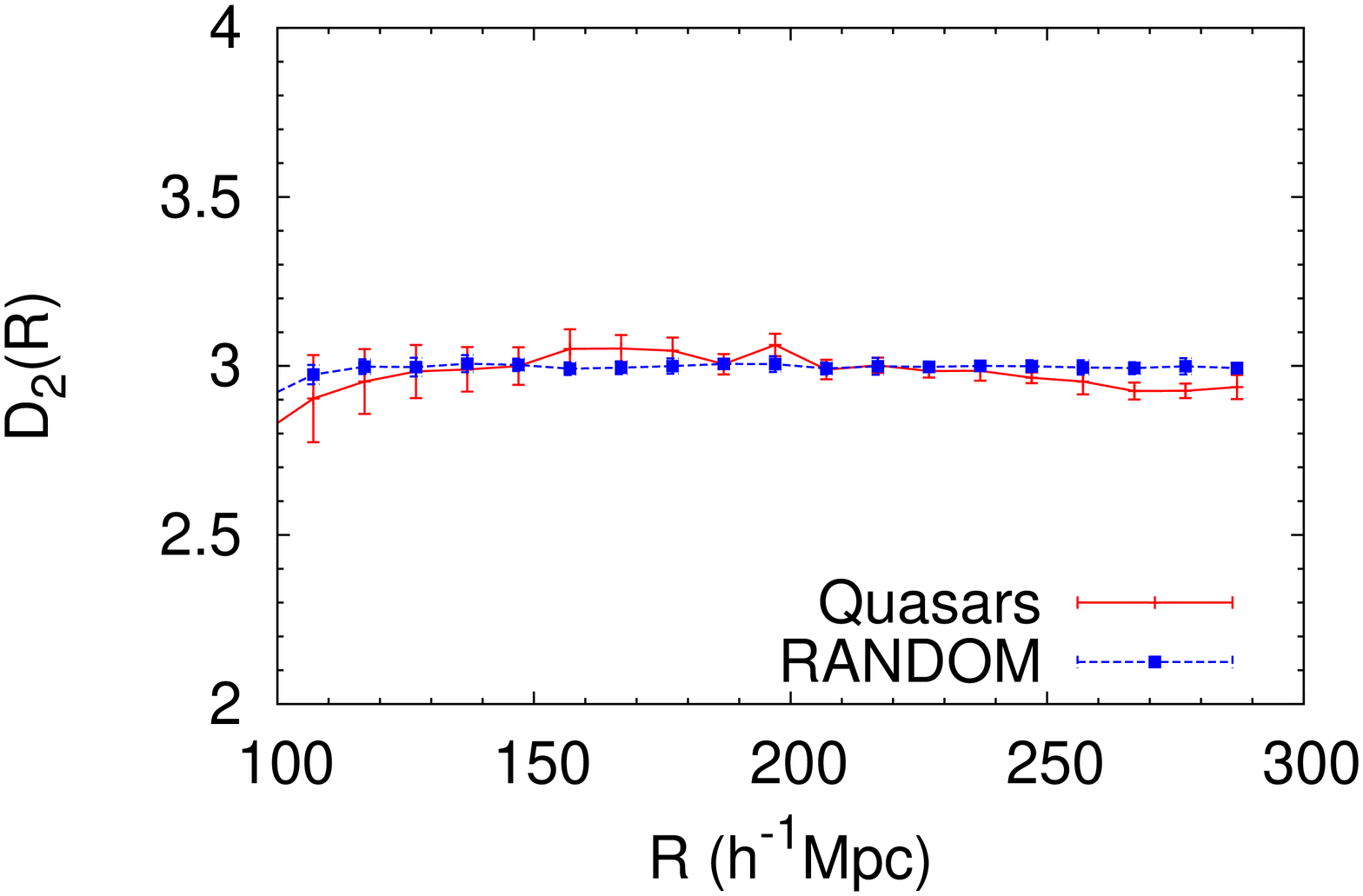}\label{fig:qso_m2}
  \includegraphics[height=0.45\textwidth, angle=-90]{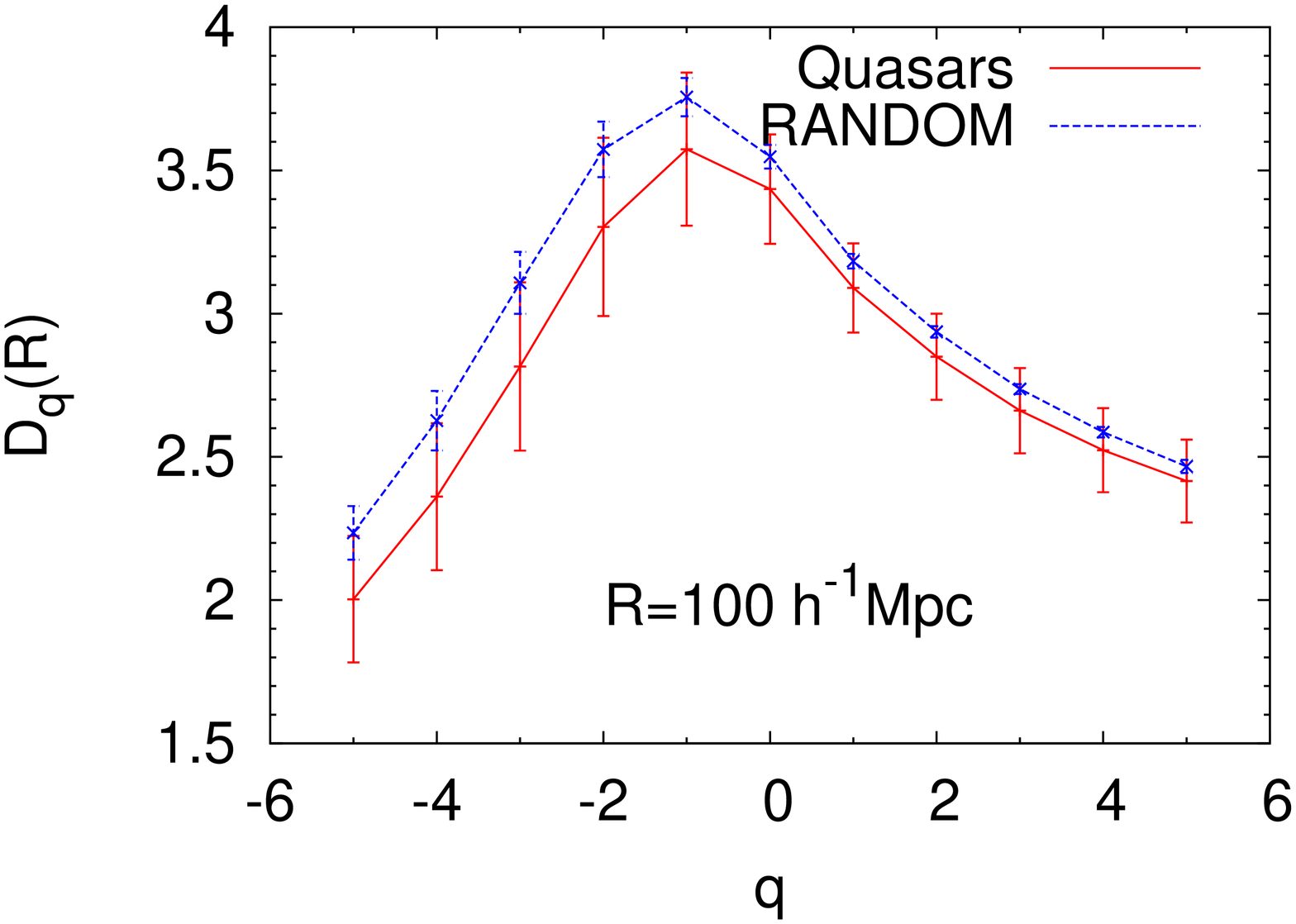}\label{fig:qso_b1}
  \includegraphics[height=0.45\textwidth,angle=-90]{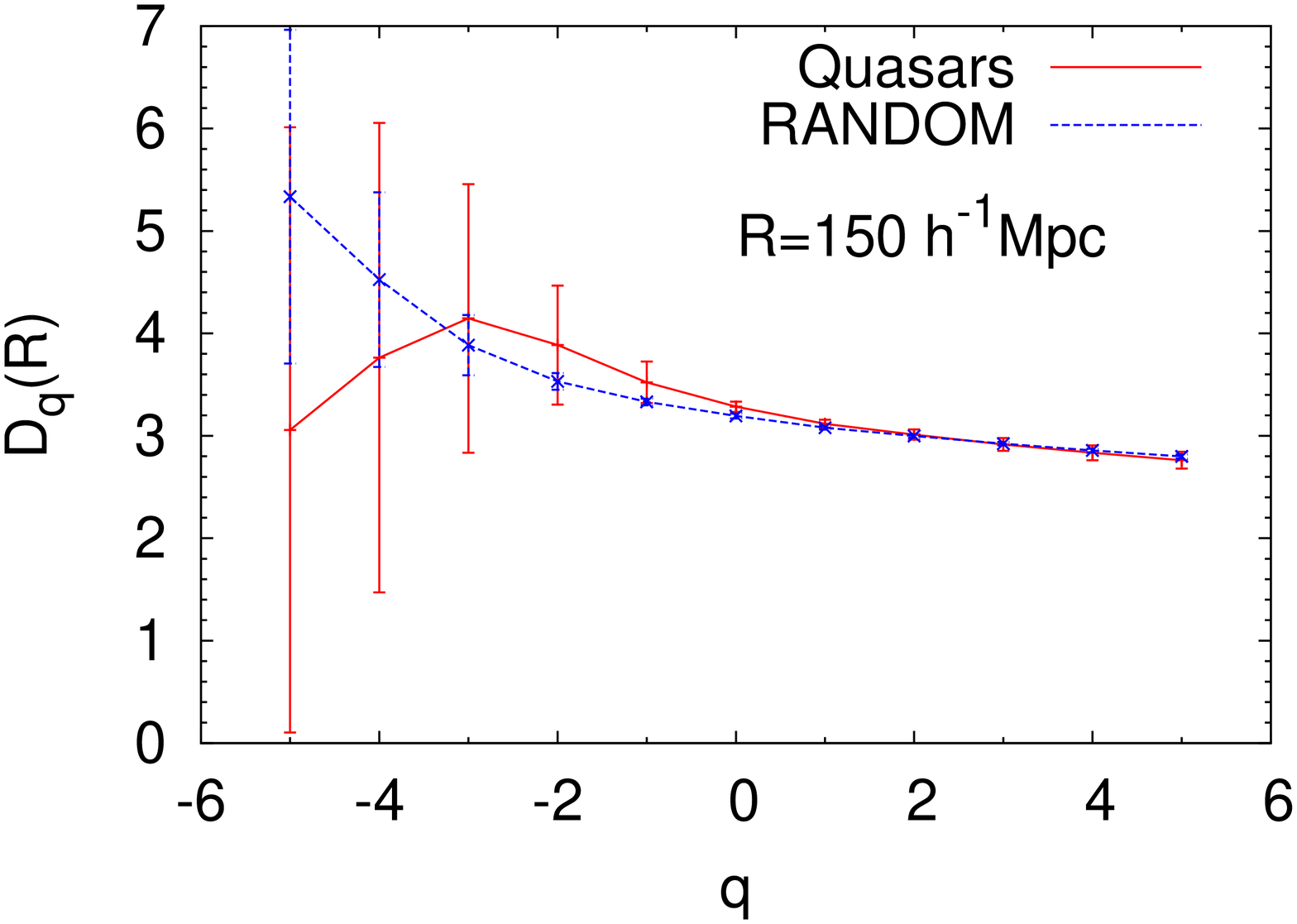}\label{fig:qso_b2}
\caption{ Same as Figure 2. but for the SDSS quasar sample.
}
\label{fig:qso_all}
\end{figure*}

\subsection{Random samples}

We have generated random samples for each of our data sets. The random
distributions have the same number density and they are distributed
over the same region having identical geometry as the actual
samples. In each case we generate $50$ independent realizations of
random samples. The random samples were analyzed in the same way as
the actual samples.

\section{Method of Analysis} 

We have used two statistical measures to study the transition to
homogeneity. One is based on the simple number counts in spheres of
radius $R$ and its scaling with $R$. The other one is the Multifractal
analysis which uses the scaling of different moments of the number
counts.

\subsection{Scaled Number Counts}

One of the simple test of homogeneity is to check how the number of
points included inside spheres changes with the radius of the
spheres. For this, we consider the galaxies as centres and place
spheres of radius $R$ around these centres to count the number of
galaxy $N(<R)$ within these spheres. We consider only those galaxies
as centres for which the spheres lie completely within the survey
region. The number counts $N(<R)$ is expected to scale as

\begin{equation}
  N(<R) \propto R^D
\end{equation}

for a homogeneous distribution, where $D$ is the ambient dimension. In
three dimension we expect $D=3$ at the scale of homogeneity. Random
distributions are considered to be homogeneous by construction. We
take the average over all the galaxies, to obtain the mean $N(<R)$ for
a distribution and define an estimator,
\begin{equation}
  \mathcal{N}(<R) = \frac{\sum\limits\limits_{i=1}^{N_c} {\rho_i}^2 \left(
    \frac{N_A(<R)}{{N_R}_i(<R)} \right)}{\sum\limits_{i=1}^{N_c}
    {\rho_i}^2},
\end{equation}
where $N_A(<R)$ and ${N_R}_i(<R)$ are the average number counts for
SDSS galaxy sample and $i^{th}-$ random samples, $\rho_i$ is the ratio
of number of galaxies in the $i^{th}$ random sample and SDSS sample
and $N_c$ is the number of random catalogues used in the analysis. It
may be noted that the number of galaxies in the random sample and the
SDSS sample may not be identical due to the exclusion of spheres near
the survey boundary. We expect the value of $\mathcal{N}(<R)$ to be
unity at the transition scale to homogeneity.

\subsection{Multifractal Analysis}

Galaxy surveys suggest that the Universe resembles a fractal on small
scales. The fractal dimensions are commonly used to characterize
fractals. There are various ways to measure the fractal dimension
\citep{borgani95}, out of which the box counting dimension and
correlation dimensions are simple to compute in finite
distributions. The Box counting dimension and the correlation
dimension quantify different aspects of the scaling behaviour. These
represents the particular cases of the generalized Minkowski-Bouligand
dimensions $D_q$, where $q$ represents different moments of galaxy
counts. The value of $D_q$ at $q=1$ and $q=2$ corresponds to Box
counting dimension and Correlation dimension respectively. For a
fractal distribution, the values of $D_q$ will be different for
different values of $q$ where $q>0$ gives more weightage to overdense
region and $q<0$ gives more weightage to the underdense region.

The generalized correlation integral is defined as,
\begin{equation}
  C_q(R) = \frac{1}{MN} \sum_{i=1}^M\left[n_i(<R)\right]^{q-1},
\end{equation}
where $n_{i}(<R)$ is the number count within a sphere of comoving
radius $R$ centered on the $i^{th}$ galaxy, $M$ is the number of
centers chosen and $N$ is the total number of galaxies.

The generalized Minkowski-Bouligand dimension is given by,
\begin{equation}
  D_q(R) = \frac{1}{q-1} \frac{d \log C_q}{d\log r},
\end{equation}

We have considered $q$ in the range $-4\le q \le 4$. The finite number
of galaxies restrict us to consider any arbitrary large values of
$|q|$ \citep{bouchet}. For the calculation of $D_q$ from $C_q$, we
have used the numerical differentiation by riddle's method describe in
\citet{press}. The random samples are considered to be homogeneous by
construction. So to determine the scale of transition to homogeneity,
we compare the results for various galaxy samples with that from the
random samples. If the galaxy sample is homogeneous then we expect the
value of $D_q$ to be consistent with that of random samples at the
transition scale to homogeneity.

\section{Results}

\subsection{Results from the SDSS DR7 Main galaxy sample}
Figure \ref{fig:NvsRMGS} shows the variation of the scaled Number
counts $\mathcal{N}(<R)$ with $R$ for the main galaxy sample of the
SDSS. The $1-\sigma$ error bars are estimated from $30$ mocks samples.
We define the transition scale to homogeneity where the value of
$\mathcal{N}(<R)$ is within $1\%$ of its expected value $1$. The
result suggests that the MAIN galaxy sample is homogeneous beyond a
length-scale of $\sim 70-80$ $\hmpc$.

We show the correlation integral $C_q(R)$ as a function of radius $R$
for $q=-2$ and $q=2$ in the top two panels of Figure \ref{fig:mgs_all}
. The solid line and the dot dashed line shows the results for the
SDSS main galaxy sample and its mock counterparts from the LasDamas
N-body simulations and they are nearly indistinguishable from each
other. The result for the random samples is shown with the dashed
line. We find that $C_q(R)$ increases with $R$ for positive values of
$q$ whereas it falls progressively for negative values of $q$. It is
quite clear from the plot that on large scales the main galaxy sample
of SDSS, the corresponding mock samples and the random samples all
exhibit the same scaling behaviour. But the scaling behaviours for the
random samples are noticeably different from that of the SDSS main
galaxy sample and the mock samples at $R<40 \hmpc$. It may be noted
that for both $q=-2$ and $q=2$ these differences diminish with
increasing length scales $R$. The behaviour of $C_{q}(R)$ is similar
for other positive and negative values of $q$.

We show $D_{q}(R)$ as a function of $R$ for $q=-2$ and $q=2$ for the
main galaxy sample in the middle two panels of Figure
\ref{fig:mgs_all}. The $1-\sigma$ error bars for the SDSS data are
obtained from $30$ mock samples. We expect $D_{q}(R)$ to match with
the ambient dimension for a homogeneous distribution. The random
samples are homogeneous by construction and we expect them to have
$D_{q}(R)=3$ at all $R$. We can see in both the middle panels of
Figure \ref{fig:mgs_all} that for the SDSS main galaxy sample and
their mock samples $D_{q}(R)$ values converge to $3$ within $1\%$ at
scales $\sim 70-80 \hmpc$. Here we adopt a working definition for the
scale of homogeneity to be the scale where $D_{q}(R)$ comes within
$1\%$ of $3$ \citep{scrim} or $1\%$ of the results of the random
samples whichever earlier. It may be noted that $D_{q}$ values for the
SDSS and random samples differs greatly on small scales despite having
the same mean inter particle separation. This clearly indicates that
the SDSS main galaxy sample is homogeneous above a length scales of
$80 \hmpc$.

In the left and right bottom panels of Figure \ref{fig:mgs_all}, we
show the variation of $D_{q}$ across the different $q$ values for
$R=80 \hmpc$ and $R=90 \hmpc$. Clearly at $R=80 \hmpc$, the $D_{q}$
values for the main galaxy sample and its mock counterparts show a
large offset from that observed in the random samples for the entire
range of positive $q$. Interestingly, increasing $R$ to $90 \hmpc$
reduces these offsets and the results for the random samples comes
well within the $1-\sigma$ errorbars of the results from the main
galaxy sample. This indicates the presence of a transition scale to
homogeneity in the SDSS main galaxy distribution at a length scale of
$90 \hmpc$.

\subsection{Results from the SDSS DR7 LRG sample}

Figure \ref{fig:NvsRLRG} shows the variation of measured Number counts
$\mathcal{N(<R)}$ with R for the SDSS LRG sample. The $1-\sigma$ error
bars are estimated from $40$ mocks samples. Following the definition
stated earlier, we find that the LRG sample becomes homogeneous at
length-scale of $\sim 150 \hmpc$.

The top two panels of Figure \ref{fig:lrg_all} show the variation of
$C_q(<R)$ with respect to $R$ for $q=-2$ and $q=2$. We find that for
$q>0$, $C_q(<R)$ increases with $R$ whereas for $q<0$, it falls
progressively. It is quite clear from the plots that the results from
the mock samples are quite consistent with the LRG data on large
scales. But there behaviours are quite different as compared to the
random samples at $r<80 \hmpc$.

In the middle two panels of Figure \ref{fig:lrg_all} we show the
variation of $D_q(R)$ with $R$ for $q=-2$ and $q=2$ for the SDSS LRG
sample.  The $1-\sigma$ error bars for the LRG sample is estimated
from $40$ Mock samples. The $D_{q}(R)$ values for the LRG sample and
its mock counterparts are quite different than the random samples at
scale $<230 \hmpc$ for $q=-2$ and $<150 \hmpc$ for $q=2$. Following
the same working definition mentioned earlier we find that the
$D_{q}(R)$ values for the LRG sample comes within $1\%$ of $3$ at a
scale $\sim 200 \hmpc$.

The variation of $D_q$ with $q$ for two different values of $R$ are
shown in the bottom two panels of Figure \ref{fig:lrg_all}. The bottom
left panel in this figure corresponds to $R=220\hmpc$. This plot
clearly shows the transition to homogeneity is reached for the
positive values of $q$ whereas the $D_q$ for negative values of $q$
still differs from the random samples indicating the non uniformity
across the underdense regions at that scale. The right panel shows the
same results but for $R=230\hmpc$. We find that the differences in the
$D_{q}$ values at negative $q$ decreases noticeably and the errorbars
for the LRG and random samples now overlap with each other. Clearly
the spectrum of $D_{q}$ values for the LRG and random data agree well
when we change $R$ from $220 \hmpc$ to $230 \hmpc$. This indicates a
transition scale to homogeneity in the LRG data at $\sim 230 \hmpc$.

\subsection{Results from the SDSS DR12 Quasar sample}

In Figure \ref{fig:NvsRQSO} we show the scaled number counts
$\mathcal{N}(<R)$ as a function of $R$ for the SDSS quasar sample. The
$1-\sigma$ error bars shown here are estimated from samples obtained
by bootstrap resampling. We note that the scaled number counts for the
quasar data does not show a distinct transition to homogeneity as seen
earlier in the main galaxy sample and the LRG sample. The quasar
distribution seems to be quite consistent with a homogeneous
distribution from small scales upto a length scale of $\sim 70
\hmpc$. But the scaled number counts show a small and near constant
deviation from its expected value $1$ beyond $70 \hmpc$. This can not
be emphasized due to a large mean inter-particle separation $\sim 91
\hmpc$ of the SDSS quasar sample. This departure does not necessarily
imply that the quasar distribution is inhomogeneous and one needs to
apply other statistical measures to asses its significance.

We show the variation of $C_q(<R)$ with $R$ for the SDSS quasars for
two different values of $q$ in the top two panels of Figure
\ref{fig:qso_all}. We find that for both $q=-2$ and $q=2$, apparently
the value of the correlation integral $C_q(<R)$ for the quasar sample
is quite close to that for the random samples throughout the entire
length scale with small differences.

In the middle left panel and right panels of Figure \ref{fig:qso_all}
we show the variation of $D_{q}$ with $R$ for the quasar sample for
$q=-2$ and $q=2$ respectively. We find that the $D_{q}$ values for the
quasar sample for $q=2$ are quite consistent with that from a random
distribution for $R>100 \hmpc$ whereas the $D_{q}$ values differ from
that a random distribution for $q=-2$ upto a length scales of $\sim
200 \hmpc$.

We show how $D_{q}(R)$ depend on $q$ for the quasar sample for $R=100
\hmpc$ and $R=150 \hmpc$ in the bottom left and right panels of figure
\ref{fig:qso_all} respectively. Clearly at $R=100 \hmpc$ the
transition to homogeneity is observed for the positive values of $q$
whereas the $D_q$ for negative values of $q$ still differs from the
random samples indicating the non uniformity across the underdense
regions at that scale. We also note that the $D_{q}$ values for both
the quasar sample and the random samples deviate from $3$ for almost
all the values of $q$. This may not be surprising as this length scale
is very close to the mean interparticle separation in the quasar and
random samples. The right panel shows the same results but for $R=150
\hmpc$. We find that the differences in the $D_{q}$ values at negative
$q$ decreases noticeably and the results for the random samples are
well within $1-\sigma$ errorbars for the results from the
quasars. Further at $R=150 \hmpc$ the $D_{q}$ values converge to $\sim
3$ for positive $q$ values whereas the deviations still persists at
negative $q$ values. Interestingly the deviations are of opposite sign
for the quasar and the random samples. \citet{roberts} find that the
generalized dimensions $D_{q}$ for $q\leq 0$ are extremely sensitive
to the regions of low density for a finite size data sets. This
implies that one cannot tell the difference between empty space and
space that should be filled in with very low probability. These
differences dramatically affect the generalized dimensions $D_{q}$ for
the negative $q$ values. Considering the very low density of the
quasar sample we conclude that the SDSS quasar distribution indicates
a transition to homogeneity at $\sim 150 \hmpc$.

\section{Conclusions}

We used the scaled number counts and different moments of the number
counts in spheres to identify the transition scale to homogeneity in
the SDSS main galaxy sample, LRG sample and quasar sample. The scaled
number counts in the main galaxy sample and LRG sample show a
transition to homogeneity on scales of $80 \hmpc$ and $150 \hmpc$
respectively whereas the scaled number counts in the quasar sample is
consistent with homogeneity on small scales with a small and near
constant deviation from it on larger scales. It may be noted here that
the quasar sample is the largest but sparsest among all the samples we
have analyzed. Further we could not define a strictly volume limited
sample for the quasars as done for the main galaxy sample. The LRG
sample is a quasi volume limited in nature. To define our quasar
sample we used a redshift dependent magnitude cut so as to maintain a
near uniform comoving number density for them. This helps us to
maintain a uniform comoving number density in the quasar sample but
introduces a non-uniformity in the magnitudes of the quasars in the
sample. Further the resulting quasar sample has a very poor number
density with mean inter-particle separation of $\sim 91 \hmpc$ and the
number counts are expected to be dominated by shot noise below this
length scale. Given the incompleteness of the quasar sample it is hard
to interpret the behaviour of the scaled number counts in the quasar
sample in a simple manner.

We analyzed the spectrum of the generalized dimension for all three
samples to find that the generalized dimension $D_{q}$ show a
transition to homogeneity at $90 \hmpc$, $230 \hmpc$ and $150 \hmpc$
for the SDSS main galaxy sample, the LRG sample and the quasar sample
respectively. These values are somewhat different than that obtained
with the scaled number counts of the respective samples. This may
arise due to higher sensitivity of $D_{q}$ to underdense regions. One
may note that these samples have quite different number densities. The
number density decreases by order of magnitudes as we go from main
galaxy sample to the LRG sample and the quasar sample. The transition
scale to homogeneity are particularly different for the LRG and the
quasar sample when they are determined from the $q$ dependence of the
generalized dimension $D_{q}$ instead of the behaviour of their scaled
number counts. It is thus important to note that the transition scale
to homogeneity is also sensitive to the statistical measures employed.

The present analysis indicates that there exist a transition scale to
homogeneity for all the tracers of the mass distribution used in this
analysis. We find the transition scale to homogeneity to be quite
different for the normal galaxies, LRGs and quasars. This may
originate from the fact that these distributions are biased tracers of
the underlying mass distribution and the large scale linear bias for
these samples are expected to be quite different from each other. The
quasars are found to inhabit dark matter halos of constant mass $\sim
2 \times 10^{12} h^{-1} M_{\odot}$ from the time of peak quasar
activity ($z \sim 2.5$) and their large scale linear bias evolves from
$b=3$ at $z\sim 2.2$ to $b=1.38$ at $z\sim 0.5$ \citep{shen, ross,
  geach}. Our quasar sample extends from $z=2.2$ to $z=3.2$ for which
we expect a large scale linear bias of $>3$. On the other hand the
SDSS LRG sample is known to have a large scale linear bias values of
$b\sim 2$ \citep{martin,sawang}. So one would expect the quasar sample
to be homogeneous on larger scales than the LRG sample. \citet{sarkar}
applied an information theory based method \citep{pandey13} to analyze
the SDSS DR12 quasar sample and find that the quasar distribution is
homogeneous beyond a length scale of $250 \hmpc$. Information entropy
is related to the higher order moments of a distribution and may be
more sensitive to the homogeneities present in a distribution. But the
present analysis shows that the SDSS quasar sample is homogeneous on a
length scale which is smaller than this and also smaller than the
transition scale to homogeneity for the LRGs. This indicates that the
clustering strength as indicated by their bias values may not be the
only parameter deciding the transition scale to homogeneity. It should
be also noted here that the quasar sample is very sparse. The poor
number density of the quasar sample together with its somewhat
incomplete nature may also yield such differences in our
results. However despite these differences all the galaxy samples from
the SDSS including the main sample, the LRG sample and the quasar
sample exhibit a transition to homogeneity on large scales. The
present analysis thus validates the fundamental assumption of cosmic
homogeneity and reaffirms that the Universe is indeed statistically
homogeneous on sufficiently large scales irrespective of the tracers.

\section{ACKNOWLEDGEMENT}
B.P. would like to acknowledge financial support from the SERB, DST,
Government of India through the project EMR/2015/001037. BP would also
like to acknowledge CTS, IIT Kharagpur, IUCAA, Pune and TIFR, Mumbai
for providing support for visit and use of its facilities for the
present work. BP thanks Somnath Bharadwaj for useful comments and
discussions.

\bsp	
\label{lastpage}

\end{document}